\documentstyle[12pt,epsf]{article}

\font\ma=cmssbx10 at 10pt
\font\mb=cmssbx10 at 12pt
\font\mc=cmbxsl10 at 10pt
\def\myref#1{$^{\rm (#1)}$}
\def\rpages#1#2#3#4{{\bf #1}:#3 (#2)}
\def\jour#1{{\it #1}}
\catcode`\@=11
\def\@sect#1#2#3#4#5#6[#7]#8{\ifnum #2>\c@secnumdepth \def\@svsec{}\else
 \refstepcounter{#1}\edef\@svsec{\csname the#1\endcsname.\ }\fi
 \@tempskipa #5\relax \ifdim \@tempskipa>\z@ \begingroup #6\relax
 \@hangfrom{\hskip #3\relax\@svsec}{\interlinepenalty \@M #8\par} \endgroup
 \csname #1mark\endcsname{#7}\addcontentsline
 {toc}{#1}{\ifnum #2>\c@secnumdepth \else
 \protect\numberline{\csname the#1\endcsname}\fi #7}\else
 \def\@svsechd{#6\hskip #3\relax \@svsec #8\csname #1mark\endcsname
 {#7}\addcontentsline {toc}{#1}{\ifnum #2>\c@secnumdepth \else
 \protect\numberline{\csname the#1\endcsname}\fi #7}}\fi \@xsect{#5}}
\def\Section{\@startsection {section}{1}{\z@}{-3.5ex plus -1ex minus
 -.2ex}{2.3ex plus .2ex}{\normalsize\bf}}
\def\Subsection{\@startsection{subsection}{2}{\z@}{-3.25ex plus -1ex minus
 -.2ex}{1.5ex plus .2ex}{\normalsize\bf}}
\def\Subsubsection{\@startsection{subsubsection}{3}{\z@}{-3.25ex plus
 -1ex minus -.2ex}{1.5ex plus .2ex}{\normalsize\bf}}
\def\section{\@startsection {section}{1}{\z@}{-3.5ex plus -1ex minus
 -.2ex}{2.3ex plus .2ex}{\normalsize\mb}}
\catcode`\@=12

\newcommand{\centre}[2]{\multicolumn{#1}{c}{#2}} 
\newcommand{\crule}[1]{\multispan{#1}{\hrulefill}} 
\newlength{\txa}
\begin{document} %
\vglue-20mm{\footnotesize\it\noindent{}Journal of Statistical Physics, Vol. 86,
Nos. 5/6, 1997\hfil} 
\Section*{\Large\bf Corner Exponents in the\\ Two-Dimensional Potts
Model\hfil} \subsubsection*{Dragi Karevski\footnote{Laboratoire de Physique du
solide (URA CNRS No 155), Universit\'e Henri Poincar\'e (Nancy~I), F-54506 
Vand\oe uvre l\`es Nancy Cedex, France.}, Peter Lajk\'o$^{1,}$\footnote{Permanent
address: Department of Theoretical Physics, University of Szeged H--6720
Szeged, Hungary.} and~Lo\"\i c~Turban$^1$
\hfil} \vglue 0pt
{\leftskip=5em\par\noindent\footnotesize
{\it Received May 14, 1995; final July 16, 1996}
\par\noindent\rule{\txa}{.25mm}\vglue 5pt
\par\noindent
The critical behavior at a corner in two-dimensional Ising and three-state Potts
models is studied numerically on the square lattice using transfer operator
techniques. The local critical exponents for the magnetization and the energy
density for various opening angles are deduced from finite-size scaling results
at the critical point for isotropic or anisotropic couplings. The
scaling dimensions compare quite well with the values expected from conformal
invariance, provided the opening angle is replaced by an effective one in
anisotropic systems.
\par\noindent\rule{\txa}{.25mm}\vglue 5pt
\par\noindent{{\ma KEY WORDS:}  Potts model; corner exponents; anisotropy; length
rescaling.} \par\leftskip=0pt}
\vglue 0pt
\section{INTRODUCTION\label{s1}}
The surface shape of a system may have some influence on its local
critical behavior at a second order phase transition. This was shown
by Cardy\myref{1} for a magnetic system with $O(N)$ symmetry within
mean-field theory and in $d=4-\varepsilon$ dimensions. The local magnetic
exponent at a wedge was found to vary continuously with the opening angle
$\theta$.

Marginal behavior was also obtained at a corner in the
two-dimensional Ising model using the star--triangle recursion relation
on the triangular lattice and the corner-to-corner spin correlation function
on the square lattice to calculate the corner
magnetization.\myref{2,\,3} The same problem was later studied
on the square lattice using row-to-row\myref{4} or corner\myref{5} transfer
matrix techniques. An expression for the $90^\circ$ corner magnetization was
conjectured in ref. 4. Recently, Abraham and Latr\'emoli\`ere\myref{6,\,7}
obtained the edge magnetization analytically, as a function of the distance from
the corner, thus confirming the conjecture of Kaiser and Peschel. 

Other systems were also considered, such as the planar
Heisenberg anti\-ferro\-ma\-gnet\myref{8} and the self-avoiding walk
confined into a wedge in two and three
dimensions.\myref{9\mbox{--}13}    

In two dimensions a varying corner exponent $x^c(\theta)$ immediately follows
from the conformal mapping $w=z^{\theta/\pi}$ which transforms the half-plane
into a wedge with opening angle~$\theta$.\myref{14,\,2} The decay of
the critical corner-to-bulk correlation functions then gives 
\begin{equation}
x^c(\theta)={\pi\over\theta}\, x^s
\label{e1.1}
\end{equation}
where $x^s$ is the corresponding surface exponent. This result is valid for
isotropic systems only. When the couplings are anisotropic, lengths have to be
rescaled in order to restore isotropy\myref{1,\,2} and~(\ref{e1.1})
still applies with an effective value of the opening angle.

The marginal local critical behavior may be understood by considering a system
with a ``parabolic" shape, $y=\pm Cx^\alpha$.\myref{15,\,16} Under a
length rescaling by a factor $b$, $C$ transforms into $C'=b^{\alpha-1}C$. Thus
the flow is towards the flat surface geometry when $\alpha>1$ and towards the
half-line when $\alpha<1$. When $\alpha=1$, i.e., for the corner geometry, the
surface shape  is scale invariant and $C=\tan(\theta/2)$ is the marginal
variable.

Up to now, the conformal result~(\ref{e1.1}) has been checked for the local
magnetic exponent~$x_m^c(\theta)$ with opening angles corresponding to simple
fractions of $\pi$. In the present work, we extend these results by
studying the local critical behavior of both the energy and the
magnetization in the two-dimensional $q$-state Potts model with $q=2,3$.
Various rational values of $\tan(\theta/2)$ and $\tan\theta$ are considered.
In Section~\ref{s2} we use a local operator formalism for the
construction of the row-to-row transfer matrices which are needed for the corner
geometry on a square lattice.  The corner exponents are deduced from a
finite-size scaling analysis of the data for isotropic and anisotropic
systems in Section~\ref{s3}.

\setcounter{equation}{0}
\section{POTTS MODEL IN THE CORNER GEOMETRY\label{s2}}
\ \ \ \ \  We consider the zero-field Potts model with Hamiltonian
\begin{equation}
-\beta{\cal H}=\sum_{i,\, j}[K_1\,(q\delta_{\sigma(i,\, j),\sigma(i+1,\, j)}-1)
+K_2\,(q\delta_{\sigma(i,\, j),\sigma(i,\, j+1)}-1)]
\label{e2.1}  
\end{equation}
where $\delta$ is the Kronecker delta function and the Potts variables
$\sigma(i,\, j)$ take the values \mbox{$0, 1,\cdots, q-1$}. When $q=2$, the Ising
model is recovered. The sum runs over the bonds of a square lattice in the corner
geometry as shown in Fig.~1. The couplings are assumed to be anisotropic
with values $K_1$ in the vertical direction and $K_2$ in the horizontal one. 

\begin{figure}
\epsfxsize=7.5cm
\begin{center}
\vglue-1.5cm
\mbox{\epsfbox{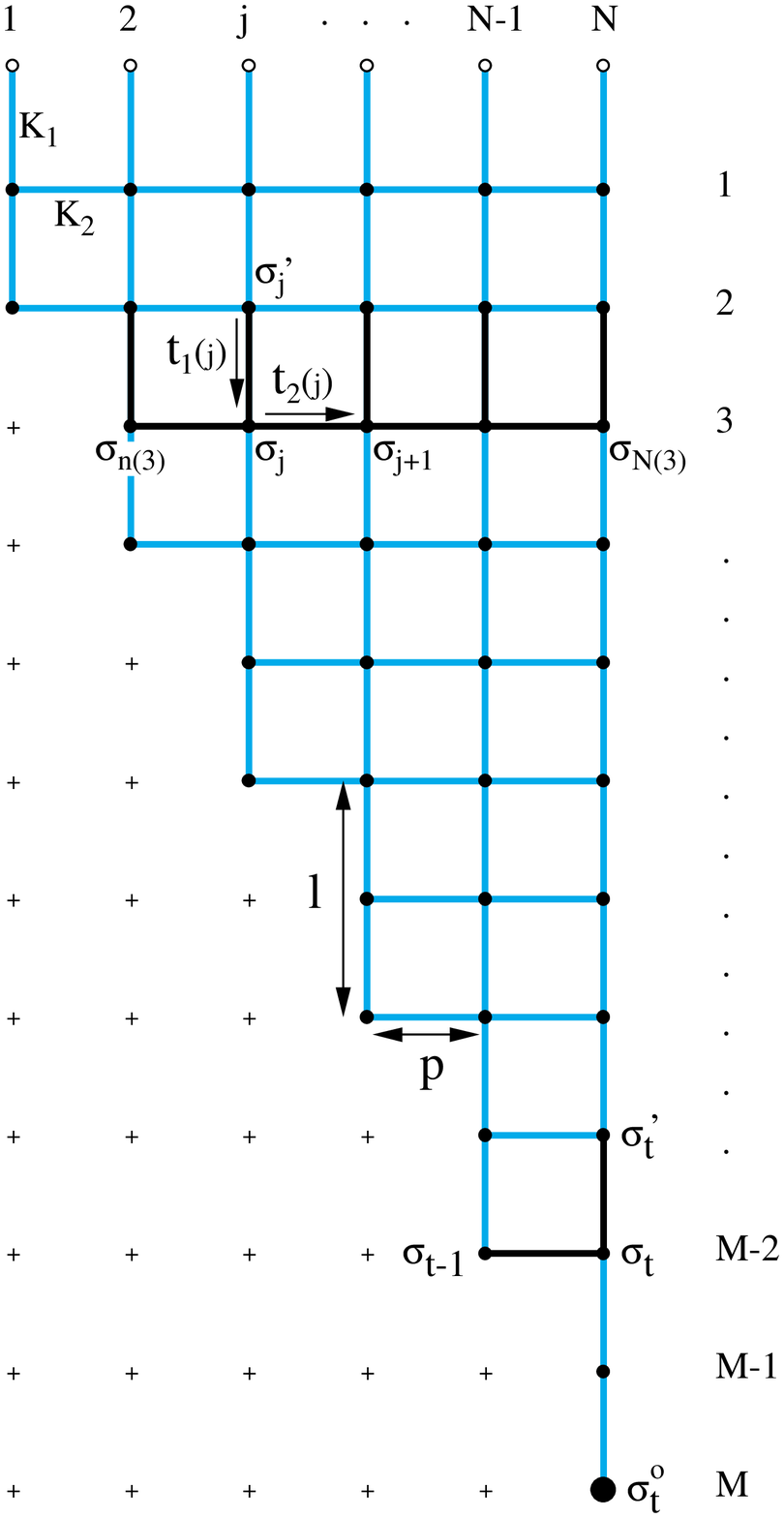}}
\vglue.5cm{\footnotesize Fig.$\!$ 1. Corner geometry and
transfer operator.}
\end{center}
\end{figure}

Let us define the row-to-row transfer matrix elements as
\begin{eqnarray}
T_{m,\, m-1}&=&\langle m\vert T_m\vert m-1\rangle\nonumber\\
&=&\exp\left[K_1\sum_{j=n(m)}^{N(m)}(q\delta_{\sigma_j,\,\sigma'_j}-1)
+K_2\sum_{j=n(m)}^{N(m)-1}(q\delta_{\sigma_j,\,\sigma_{j+1}}-1)\right]
\label{e2.2}
\end{eqnarray}
where $\sigma_j$ ($\sigma'_j$) denote the Potts variables on the $m$th
[$(m-1)$th] row, respectively. $\vert m\rangle$ is a state vector corresponding
to a configuration of the Potts variables from $n(m)$ to $N(m)$ along the $m$th
row and $T_m$ is the transfer operator acting on this state vector.

The transfer operator can be written as the product $T_m=V_2(m)V_1(m)$
with~\myref{17} 
\begin{eqnarray}
V_1(m)&=&\exp\left[K_1^*\sum_{j=n(m)}^{N(m)}\sum_{p=1}^{q-1}(R_j)^p\right]
\nonumber\\
V_2(m)&=&\exp\left[K_2\sum_{j=n(m)}^{N(m)-1}\sum_{p=1}^{q-1}(C_j^\dagger
C_{j+1})^p\right]
\label{e2.3}
\end{eqnarray}
$K_1^*$ is a dual coupling such that $x_1\, x_1^*=q$
with \mbox{$x_i=\exp(qK_i)-1$}.
The operator $C_j$ is diagonal on the basis of the Potts states
$\vert\sigma_j\rangle$ whereas $R_j$ acts as a ladder operator on the same basis,
so that
\begin{equation}
C_j\vert\sigma_j\rangle=\exp\Big(i{2\pi\over
q}\,\sigma_j\Big)\,\vert\sigma_j\rangle\, ,\qquad
(R_j)^p\vert\sigma_j\rangle=\vert\sigma_j+p\rangle\ \pmod{q}
\label{e2.5}
\end{equation}

Introducing local transfer operators
\begin{equation}
t_1(j)=x_1+\sum_{p=0}^{q-1}(R_j)^p\, ,\qquad t_2(j)=1+{x_2\over
q}\sum_{p=0}^{q-1}(C_j^\dagger C_{j+1})^p
\label{e2.6}
\end{equation}
we have that, up to an unimportant constant factor, the
row-to-row transfer operator appearing in~(\ref{e2.2}) can be rewritten as
\begin{equation}
T_m=\prod_{j=n(m)}^{N(m)-1}t_2(j)\,\prod_{j=n(m)}^{N(m)}t_1(j)
\label{e2.7}
\end{equation}

This transfer operator may be viewed as a discrete time evolution operator
acting on the Potts state vector at time $m-1$ to make it evolve to its state at
time $m$. Due to the corner geometry, this operator is time-dependent.
Given an initial state $\vert i\rangle$ at time 0, corresponding to the top
edge in~Fig.~1, this state will evolve to 
\begin{equation}
\vert m\rangle=T_m\, T_{m-1}\cdots T_2\, T_1\vert i\rangle
\label{e2.8}
\end{equation}
at time $m$. In the sequel we shall use either free or fixed boundary
conditions on the top edge. Fixed boundary conditions correspond to an initial
state \mbox{$\vert k\rangle=\vert\sigma_1\sigma_2\cdots\sigma_N\rangle$} whereas
for free boundary conditions one has to take a superposition 
$
\vert f\rangle=q^{-N/2}\sum_{k=1}^{q^N}\vert k\rangle
$
of the $q^N$ configurations of the $N$ Potts variables with equal amplitudes. 
One may notice in~Fig.~1 that a site variable $\sigma_j$ with $j<n(m)$ or
$j>N(m)$ remains frozen at time $m$ and later.

The corner magnetization $m_c$ is calculated with the Potts
variable $\sigma_t^0$, where $t$ is the index of the column
corresponding to the tip (see~Fig.~1). It is given by\myref{18}
\begin{equation}
m_c={q\langle\delta_{\sigma_t^0,0}\rangle-1\over q-1}\, ,\qquad
\langle\delta_{\sigma_t^0,0}\rangle={1\over q}\sum_{p=0}^{q-1}{\langle
f\vert C_t^p\vert M\rangle\over\langle f\vert M\rangle}
\label{e2.11}
\end{equation}
where the average of the Kronecker delta follows from~(\ref{e2.5}). The state
$\vert f\rangle$ corresponds to free boundary conditions on the sides of the
wedge whereas  $\vert M\rangle$ is defined as in~(\ref{e2.8}) with \mbox{$\vert
i\rangle=\vert00\cdots0\rangle$}. Thus the top edge is fixed with all the sites
in the state 0. At the critical point this symmetry-breaking boundary condition
ensures that the tip magnetization remains non-vanishing on a finite system. 

The energy density associated with a bond can be defined, up to a constant
factor, as the average of the Kronecker delta. When the bond is horizontal,
the system evolves as above up to time \mbox{$M-2$}, where the value
of $\delta_{\sigma_{t-1},\sigma_t}$ is taken (see~Fig.~1), and the corner
energy density is given by 
\begin{equation}
e_c^h=\langle\delta_{\sigma_{t-1},\sigma_t}\rangle={1\over
q}\sum_{p=0}^{q-1}{\langle f\vert T_MT_{M-1}(C_{t-1}^\dagger C_t)^p\vert
M-2\rangle \over\langle f\vert M\rangle} 
\label{e2.12}
\end{equation}
With a vertical bond one obtains nonvanishing contributions to the average of
the Kronecker delta when there is no flip along this bond, so that $t_1(t)$
in~(\ref{e2.6}) contributes a factor \mbox{$1+x_1$}. With the geometry
of~Fig.~1 the energy density can be written as
\begin{equation}
e_c^v=\langle\delta_{\sigma_t,\sigma'_t}\rangle=(1+x_1)\,{\langle f\vert
T_MT_{M-1}\prod_jt_2(j)\prod_{j\neq t}t_1(j)\vert M\!-\!3\rangle\over\langle
f\vert M\rangle} 
\label{e2.13}
\end{equation}
These expressions are easily generalized for other shapes. 

\setcounter{equation}{0}

\setcounter{equation}{0}
\section{NUMERICAL RESULTS\label{s3}}
\ \ \ \ \  The Potts exponents for the corner magnetization and the corner energy
density have been obtained through finite-size scaling at the critical point for
different opening angles, using the shapes sketched in~Fig.~2. In order to
limit the size $N$ of the system, we use only unit steps ($p=1$) in the
horizontal direction and change the angle $\theta$ by varying $l$
(see~Fig.~1). The maximum sizes studied are \mbox{$N=19$} for \mbox{$q=2$}
(Ising model) and \mbox{$N=11$} for \mbox{$q=3$}, so that an extrapolation of
the finite-size estimates of the exponents is needed. Isotropic and anisotropic
sytems have been considered.

\begin{figure}
\epsfxsize=7.5cm
\begin{center}
\vglue-2cm
\mbox{\epsfbox{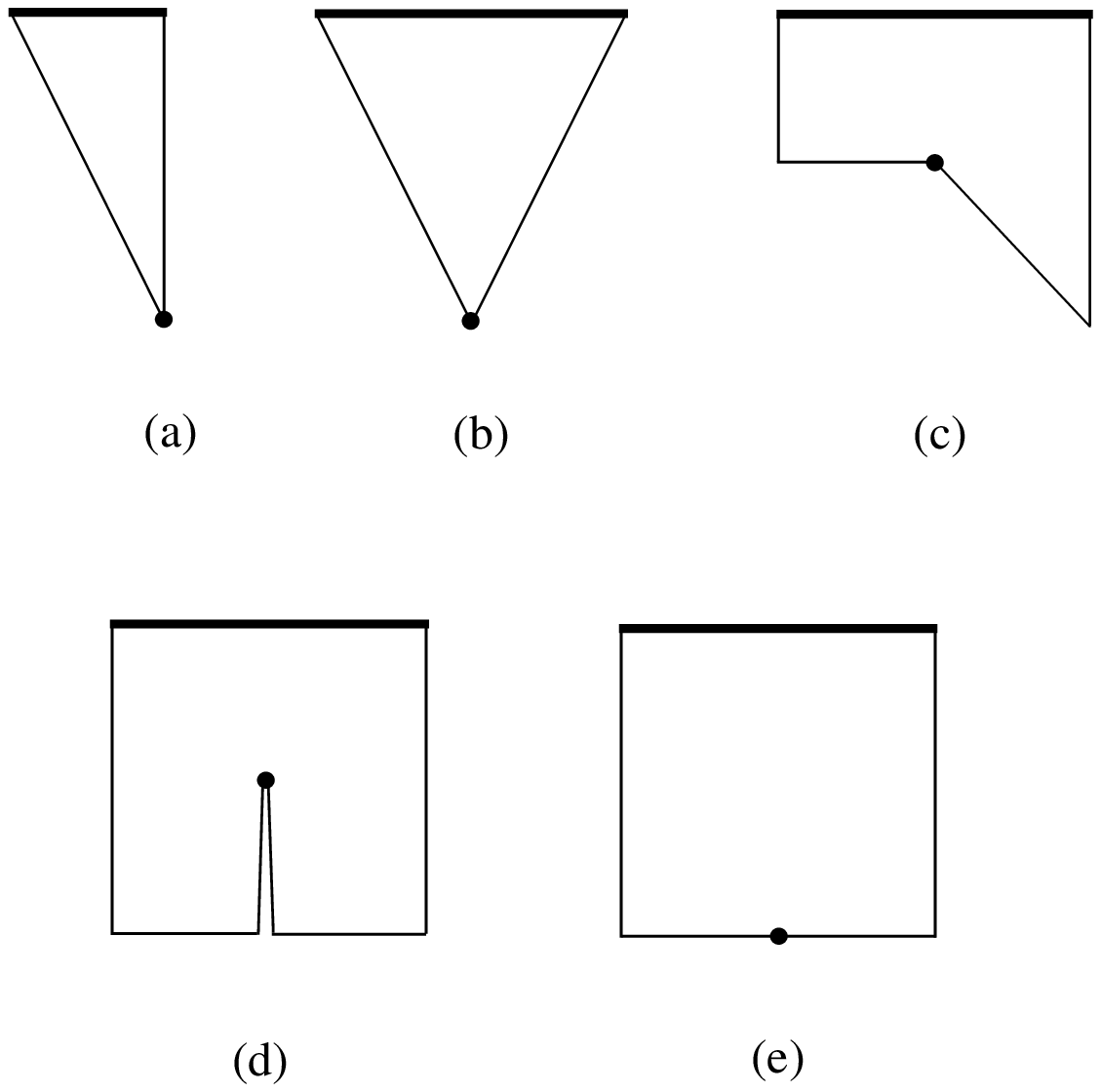}}
\vglue.5cm{\footnotesize Fig.$\!$ 2. The different system shapes used to calculate the corner magnetization
and energy density. The Potts states at the top eddge (bold line) are either
fixed or free.}
\end{center}
\end{figure}

When the system is isotropic the critical point,
corresponding to \mbox{$x_1x_2=q$}, is located at
\mbox{$K_c=(1/q)\ln(1+\sqrt{q})$}. The critical corner magnetization
decays as 
\[
m_c^{\rm crit}(N)=A_m(\theta)\, N^{-x_m^c}
\]
with the system size~$N$. 

The corner exponent $x_m^c(\theta)$ deduced from sequence extrapolations of the
finite-size estimates using the BST algorithm\myref{19} is given in
Table~I. The numerical results are in quite good agreement with the values
expected from conformal invariance in Eq. (\ref{e1.1}) with the surface magnetic
exponents at the ordinary surface transition \mbox{$x_m^s=1/2$} for \mbox{$q=2$}
and \mbox{$x_m^s=2/3$} for \mbox{$q=3$}.\myref{14,\,16} 
The critical corner energy density
\[
e_c^{\rm crit}(N)=e_c^{\rm crit}(\infty)+A_e(\theta)\, N^{-x_e^c}
\]
contains a regular part which depends on the opening angle and the bond
orientation. 
\begin{table}
\small
\begin{center}
{\ma Table $\!$I.  Scaling Dimension of the Corner
Magnetization for the {\mc q}-State Potts Model as a Function of the 
Opening Angle.$^{a}$}
\vglue4mm
\begin{tabular}{@{}*{7}{l}}
\hline\hline\hline 
&\centre{2}{$x_m^c$}&&&\centre{2}{$x_m^c$}\\ 
&\crule{2}&\
&&\crule{2}\\ 
$\theta$ (deg)&$q=2$&Expected&&$\theta$ (deg)&$q=3$&Expected\\
\hline
360(d)&0.25(1)&0.25&&180(e)&0.667(1)&0.6667\\
270(c)&0.332(1)&0.333333&&90(b)&1.331(3)&1.3333\\
225(c)&0.400(1)&0.4&&53.13(b)&2.243(3)&2.2586\\
180(e)&0.5000(1)&0.5&&45(a)&2.66(1)&2.6667\\
90(b)&0.999994(6)&1&&26.56(a)&4.516(3)&4.5172\\
53.13(b)&1.693961(7)&1.693955&&18.43(a)&6.52(2)&6.5094\\
45(a)&2.00000(2)&2&&14.04(a)&8.53(4)&8.5493\\
36.87(b)&2.44098(6)&2.441016&&11.31(a)&10.58(4)&10.6101\\
28.07(b)&3.2057(3)&3.205986&&9.46(a)&12.66(4)&12.6819\\
22.62(b)&3.979(4)&3.978804&&&&\\
11.42(b)&7.8796(6)&7.880092&&&&\\
\hline\hline\hline
\end{tabular}
\\
\medskip
{\footnotesize $^a$The numbers in parentheses give the estimated uncertainty in the last digit.
The letters refer to the system shapes in~Fig.~2.}
\end{center}
\end{table}
In order to obtain the corner exponent the regular contribution has
to be substracted. This can be done by calculating \mbox{$e_c^{\rm crit}(N)$} for
either fixed or free boundary conditions at the top edge. The asymptotic value
is the same in both cases, but the amplitudes of the finite-size corrections
are different. Taking the difference, one obtains  
\[
\Delta e_c(N)=[A_e^{\rm fixed}(\theta)-A_e^{\rm free}(\theta)]\, N^{-x_e^c}
\]
Proceeding in this way, one avoids a systematic error, linked with
the estimation of the regular part, and the finite-size corrections are amplified
because the signs of the amplitudes are different for the two boundary
conditions. 

\begin{table}
\small
\begin{center}
\medskip{\ma Table $\!$II. As in Table~I, for the Corner Energy Density.}
\vglue4mm
\begin{tabular}{@{}*{4}{l}}
\hline\hline\hline
&\centre{3}{$x_e^c$}\\
&\crule{3}\\
$\theta$ (deg)&$q=2$&q=3&Expected\\
\hline
180(e)&2.001(1)&2.01(3)&2\\
90(b)&3.999(1)&4.00(4)&4\\
45(a)&7.993(7)&8.0(1)&8\\
26.56(a)&13.52(2)&13.5(1)&13.5516\\
18.43(a)&19.6(4)&19.7(4)&19.5281\\
\hline\hline\hline
\end{tabular}
\end{center}
\end{table}

The extrapolated values of the finite-size estimates for the corner exponent
$x_e^c(\theta)$ are given in~Table~II. They do not depend on the bond
orientation, as expected, and are identical for \mbox{$q=2$} and
\mbox{$q=3$}. This is consistent with the conformal result, since, at the
ordinary transition, the surface exponent $x_e^s$ which enters in~(\ref{e1.1}) 
is equal to 2, the dimension of the system, quite generally.\myref{20,\,16} The
numerical values are in good agreement with the results expected from conformal
invariance, too.

The conformal expression for the corner exponent~(\ref{e1.1}) only applies to
systems where the correlations are isotropic. In the case of an anisotropic
system, with a coupling constant ratio \mbox{$r=K_1/K_2$}, the correlation 
lengths are different and take the values $\xi_1$ and $\xi_2$ in the vertical
and horizontal directions, respectively. Isotropy can be restored by changing
the lattice parameters $a_1$ and $a_2$ in such a way that, in the rescaled 
system, the correlation lengths become the same, i.e.,  
\mbox{$\xi_1a_1=\xi_2a_2$}.\myref{1,\,2} As a consequence, one obtains
an effective opening angle which, in the geometry of~Fig.~2a, is given by
\begin{equation}
\tan\theta_{\rm eff}=\zeta\tan\theta\, ,\qquad\zeta={a_2\over
a_1}={\xi_1\over\xi_2}
\label{e3.5}
\end{equation}
where $\zeta$ is the anisotropy factor. 

For the Potts model, the following form of the anisotropy factor at the
critical point has been conjectured:\myref{21}
\begin{equation}
\zeta=\tan{\pi u\over2\lambda}\, ,\qquad{\sin
u\over\sin(\lambda-u)}={x_{1c}\over\sqrt{q}}={\sqrt{q}\over x_{2c}}\, ,
\qquad2\cos\lambda=\sqrt{q}
\label{e3.6}
\end{equation}
When \mbox{$q=2$}, it reduces to the known exact result 
$
\zeta={\cosh 2K_{1c}/\cosh 2K_{2c}}
$
for the Ising model.\myref{2}

\begin{table}
\small
\begin{center}
\medskip{\ma Table $\!$III. Corner Exponents for the Ising Model on an
Anisotropic Lattice as a Function of the Anisotropy Ratio {\mc r}.$^a$}
\vglue4mm
\begin{tabular}{@{}*{6}{l}}  
\hline\hline\hline
&\centre{2}{$x_m^c$}&&\centre{2}{$x_e^c$}\\  &\crule{2}&\ &\crule{2}\\ 
$r$&numerical&Expected&&Numerical&Expected\\
\hline
0.200&4.5737(8)&4.57374&&18.2(6)&18.295\\
0.300&3.595(2)&3.59518&&14.36(8)&14.381\\
0.400&3.05(2)&3.06772&&12.24(8)&12.271\\
0.500&2.7326(1)&2.73264&&10.92(5)&10.931\\
0.600&2.4999(5)&2.49960&&9.96(5)&9.998\\
10.00&1.16(2)&1.16200&&4.6(1)&4.648\\
\hline\hline\hline
\end{tabular}
\\
\medskip
{\footnotesize $^a$ We used the shape of
Fig. 2a with an opening angle $\theta=45^\circ$. The numerical
values are compared to the conformal result with a rescaled angle, as explained
in the text.}
\end{center}
\end{table}
\begin{table}
\small
\begin{center}
\medskip
{\ma Table $\!$IV. As in Table III, for the Three-State Potts
Model.} 
\vglue4mm
\begin{tabular}{@{}*{6}{l}} 
\hline\hline\hline
&\centre{2}{$x_m^c$}&&\centre{2}{$x_e^c$}\\ 
&\crule{2}&\ &\crule{2}\\ 
$r$&Numerical&Expected&&Numerical&Expected\\
\hline
0.109&9.31(3)&9.30667&&27(1)&27.920\\
0.166&6.98(4)&7.00928&&21.2(9)&21.028\\
0.259&5.30(3)&5.31052&&15.9(2)&15.931\\
0.414&4.05(4)&4.05604&&12.1(1)&12.168\\
0.692&3.137(6)&3.13157&&9.3(1)&9.365\\
1.241&2.45(1)&2.45244&&7.3(1)&7.357\\
2.561&1.955(6)&1.95572&&5.8(2)&5.867\\
7.530&1.5(1)&1.59445&&4.6(2)&4.783\\
\hline\hline\hline
\end{tabular}
\end{center}
\end{table}
The corner exponents $x_m^c$ and $x_e^c$ obtained on anisotropic systems with an
opening angle \mbox{$\theta=45^\circ$} in~Fig.~2a and different anisotropy
ratio $r$ are given in Tables~III and~IV. They are in good agreement with the
conformal result~(\ref{e1.1}) when the opening angle is replaced by its
effective value given by~(\ref{e3.5}). 

\setcounter{equation}{0}
\section{CONCLUSION\label{s4}}
\ \ \ \ \  The scaling dimensions of the tip magnetization and energy density at
a corner have been calculated for the Ising and three-state Potts models in two
dimensions as functions of the opening angle $\theta$. Using transfer
operator techniques and finite-size scaling at the critical point, we have
tested the result of conformal invariance, predicting a simple relation
between corner exponents, surface exponents, and opening angle, for a broad
spectrum of $\theta$ values. In the case of the Ising model, the magnetization
results extend previous studies where simple fractions of $\pi$ were considered,
whereas the energy results are new. Isotropic and anisotropic systems have been
treated. In the later case, the conformal result still applies, provided the
opening angle is properly rescaled in order to restore isotropy. 

As a possible
extension of this work, one may mention the possiblity to consider the same
problem for noninteger values of $q$, using the relation of the Potts model with
either Withney polynomials\myref{22} or ice-rule vertex
models.\myref{23,\,24}

\section*{ACKNOWLEDGMENTS}
\ \ \ \ \  P.L. thanks the Hungarian Soros Foundation for a travel grant. This
work was supported by CNIMAT under project No 155C96 and by the Hungarian
National Research Fund under grant OTKA TO12830. 

\section*{REFERENCES}

\catcode`\@=11
\def\footnotesize{\@setsize\footnotesize{12pt}\xpt\@xpt
\abovedisplayskip 10pt plus2pt minus5pt\belowdisplayskip \abovedisplayskip
\abovedisplayshortskip \z@ plus3pt\belowdisplayshortskip 6pt plus3pt minus3pt
\def\@listi{\leftmargin\leftmargini \topsep 6pt
plus 2pt minus 2pt\parsep 0pt plus 0pt minus 0pt
\itemsep \parsep}}
\catcode`\@=12

\footnotesize
\begin{enumerate}

\item{}{J.\ L.\ Cardy, \jour{J.\ Phys.\ A} \rpages{16}{1983}{3617}{}.}
\item{}{M.\ N.\ Barber,\ I.\ Peschel,\ and P.\ A.\
Pearce, \jour{J.\ Stat.\ Phys.} \rpages{37}{1984}{497}{}.} 
\item{}{I.\ Peschel, \jour{Phys.\ Lett.} \rpages{110A}{1985}{313}{}.}
\item{}{C.\ Kaiser\ and\ I.\ Peschel, \jour{J.\ Stat.\ 
Phys.} \rpages{54}{1989}{567}{}.} 
\item{}{B.\ Davies\ and\ I.\ Peschel, \jour{J.\ Phys.\
A} \rpages{24}{1991}{1293}{}.} 
\item{}{D.\ B.\ Abraham\ and\ F.\ T.\
Latr\'emoli\`ere, \jour{Phys.\ Rev.\ E} \rpages{50}{1994}{R9}{}.} 
\item{}{D.\ B.\ Abraham\ and\ F.\ T.\
Latr\'emoli\`ere, \jour{J.\ Stat.\ Phys.} \rpages{81}{1995}{539}{}.} 
\item{}{C.\ Kaiser\ and I.\ Peschel, \jour{J.\ Phys.\ C} \rpages{6}
{1994}{1149}{}.}
\item{}{A.\ J.\ Guttmann\ and\ G.\ M.\ Torrie, \jour{J.\ Phys.\
A} \rpages{17}{1984}{3539}{}.}
\item{}{J.\  L.\ Cardy\ and\ S.\ Redner, \jour{J.\ Phys.\
A} \rpages{17}{1984}{L933}{}.} 
\item{}{B.\ Duplantier\ and\ H.\ Saleur, \jour{Phys.\ Rev.\
Lett.} \rpages{57}{1986}{3179}{}.} 
\item{}{D.\ Considine\ and\ S.\ Redner, \jour{J.\ Phys.\
A} \rpages{22}{1989}{1621}{}.} 
\item{}{C.\ Vanderzande, \jour{J.\ Phys.\
A} \rpages{23}{1990}{563}{}.}
\item{}{J.\ L.\ Cardy, \jour{Nucl.\ Phys.\
B} \rpages{240}{1984}{514}{}.} 
\item{}{I.\ Peschel,\ L.\ Turban,\ and\ F.\
Igl\'oi, \jour{J.\ Phys.\ A} \rpages{24}{1991}{L1229}{}.} 
\item{}{F.\ Igl\'oi,\ I.\ Peschel\ and\ L.\ Turban, \jour{Adv.\
Phys.} \rpages{42}{1993}{683}{}.} 
\item{}{L.\ Mittag\ and\ M.\ J.\ Stephen, \jour{J.\ Math.\
Phys.} \rpages{12}{1971}{441}{}.} 
\item{}{F.\ Y.\ Wu, \jour{Rev.\ Mod.\ Phys.} \rpages{54}{1982}{235}{}.} 
\item{}{M.\ Henkel\  and\ G.\ Sch\"utz, \jour{J.\ Phys.\
A} \rpages{21}{1988}{2617}{}.} 
\item{}{T.\ W.\ Burkhardt\ and\ J.\ L.\ Cardy, \jour{J.\
Phys.\ A} \rpages{20}{1987}{L233}{}.} 
\item{}{D.\ Kim\ and\ P.\ A.\ Pearce, \jour{J.\ Phys.\ A} \rpages{20}
{1987}{L451}{}.}
\item{}{H.\ W.\ J.\ Bl\"ote\ and M.\ P.\
Nightingale, \jour{Physica} \rpages{112A}{1982}{405}{}.} 
\item{}{H.\ N.\ V.\ Temperley\ and\ E.\ H.\ Lieb, \jour{Proc.\
R.\ Soc.\ London A} \rpages{322}{1971}{251}{}.} 
\item{}{R.\ J.\ Baxter,\ S.\
B.\ Kelland,\ and\ F.\ Y.\ Wu, \jour{J.\ Phys.\ A} \rpages{9}{1976}{397}{}.} 

\end{enumerate}
\end{document}